\documentclass[reprint,groupedaddress,showkeys,showpacs,aip,author-year]{revtex4-1}
\usepackage{graphicx}
\usepackage{dcolumn}
\usepackage{bm}%
\usepackage{amsmath}

\begin{document}

\title{Time-like detonation in presence of magnetic field}

\date{\today}

\author{Shailendra Singh}
\affiliation{Indian Institute of Science Education and Research Bhopal, Bhopal, India}

\author{Ritam Mallick} 
\email{mallick@iiserb.ac.in}
\affiliation{Indian Institute of Science Education and Research Bhopal, Bhopal, India}

\begin{abstract}
We study the effect of magnetic field in an implosion process achieved by radiation. A time-varying sinusoidal magnetic field is seen to affect the continuous transition of space-like detonation to time-like detonation at the core of implosion region.
The oscillating varying magnetic field has a significant effect in increasing the volume of the time-like detonation of the core of implosion and also modify the time of the 
implosion process. This transition can have significant outcome both theoretically and experimentally in the areas of high energy hadronization of quark-gluon plasma (QGP)
matter and Inertial Confinement Fusion efforts of fuels.
\end{abstract}



\maketitle

\section{Introduction}

The hydrodynamics of shock waves has found numerous applications in the field of fluid dynamics, astrophysics, high energy physics, and cosmology.
A shock wave is a wave where the disturbance in the medium propagates faster than the local speed of sound. At either side of the interference, the thermodynamic variables
vary discontinuously. It occurs when there is a rapid compression or expansion of the system. The mass, momentum, and energy conservation laws across the surface lead to the 
Rankine-Hugoniot (RH) and Taub equation \citep{taub,landau} connecting the properties of the fluid on the either of the discontinuity. The normal vector of the surface of discontinuity is space-like (SL), and the wave propagation velocity is less than the speed of light.

Almost four decades later it was realized \citep{csernai} that under some condition the discontinuity of the surface could also be time-like (TL) and there can be a 
rapid phase transition. Arguing that if a system undergoes a rapid rarefaction, bubbles can form at different spatial points which are causally disconnected.
If the thickness of the surface forming bubbles is very thin, the boundary between the two phases of matter becomes TL. The inflationary model of the universe was thought to be one such example. 

Such treatment was successful in describing the sudden and rapid hadronization of quark-gluon plasma in high energy physics \citep{csernai2} which would have been missed by general 
SL fronts. In such dense matter, an implosion induced by fast burning can smoothly take an SL detonation to TL detonation. TL detonation found limited use in astrophysics \citep{mallick} 
and in most practical purposes the shocks are slow.

Recently there has been experimental efforts to observe 
Inertial confinement fusion (ICF) \citep{hurricane, park,casey}. The experiments failed due to the appearance of Rayleigh-Taylor
surface instabilities. Such instabilities can be avoided if the detonation front moves with the speed of light, which in turn can be achieved by radiation \citep{csernai3}.
Therefore, such a theoretical model can have a practical application as well.

In this work, we carry forward the theoretical work by Csernai \citep{csernai} and show the effect of magnetic field on the continuous transition from SL to 
TL detonation in implosion achieved via radiation. In Section II we calculate the TL detonation front due to radiation. In section III we introduce magnetic field in our calculation. 
In section IV we present our result and finally in section V we draw our conclusion from our results.

\section{Time-like detonation due to radiation}

Here we have assumed that there is a spherical core filled with matter having vanishing opacity. 
The core is surrounded by a rapidly igniting shell whose radiation is responsible for the heating of the center.
When the temperature reaches a specific value $T_{c}$, it follows an exothermic transition. 
Neglecting the compression of the fuel, the heating of the inner core is assumed to be due to isotropic radiation.
The radius of the shell (R) remain unchanged, R = Constant. The shell is ignited at time $t_{0} = 0$ at all points simultaneously. 
$Q$ is the heat that the shell radiates in unit time through a unit surface and $\kappa$ be the fraction of heat absorbed by the matter. 
Then at a distance $r$ from the center of the core (ignoring the opacity of core and measuring the distance in units of $R$ and time in the units of $R/c$) the change
in heat per unit time is 
\begin{equation} \label{eq1}
\frac{dQ}{dt}= \kappa Q\int_{0}^{t}d\tau\int_{0}^{2\pi}d\phi\int_{1}^{0}d(\cos{\theta})\frac{\delta(\tau-|\vec{r}^{'}|)}{|\vec{r}^{'}|^2}
\end{equation}
where $r^{'2}=1+r^2-2r\cos{\theta} $, 
(only radiation that have reached inside the radius $r$ will contribute to heating of the matter). 
Assuming $R=c=1$, we have
\begin{eqnarray} \label{eq2}
\frac{dQ}{dt}= \kappa Q\int_{0}^{t}d\tau\int_{0}^{2\pi}d\phi\int_{1}^{0}d(\cos{\theta}) \\ \nonumber
\frac{\delta(\tau-(1+r^{2}-2r\cos{\theta})^{\frac{1}{2}})}{1+r^{2}-2r\cos{\theta}}.
\end{eqnarray}    
From the property of the $\delta$ function we can write $$ \delta(g(\cos{\theta}))=\frac{\delta(\cos{\theta}-\cos{\theta_0})}{g{'}(\cos{\theta_0})}$$
and $\theta_0$ is the angle for which $g(cos{\theta})$ is zero.

Defining the function $g(cos{\theta})$ as
\begin{equation}
 g(cos{\theta})=\tau-(1+r^{2}-2r\cos{\theta})^{\frac{1}{2}}, \\ \nonumber
\end{equation}
we have 
\begin{equation}
g'(\cos{\theta})= \frac{r}{(1+r^{2}-2r\cos{\theta})^{\frac{1}{2}}}. \nonumber
\end{equation}
For $ \cos{\theta_0}$, we have 
\begin{eqnarray*}
 \tau-(1+r^{2}-2r\cos{\theta_0})^{\frac{1}{2}} = 0  \\ 
 \Rightarrow \cos{\theta_0}=\frac{1+r^{2}-\tau^{2}}{2r}.
\end{eqnarray*}
Therefore, eqn \ref{eq2} becomes dependent on $c\tau$ distance covered by the radiation in the range $(1-r)$ to $(1+r)$
   $$\frac{dQ}{dt}=\frac{2\pi\kappa Q}{r}\int_{1-r}^{a}=\frac{2\pi\kappa Q}{r}[\ln{\tau}|_{1-r}^{a}], $$
   where 
  \[ a =
  \begin{cases}
  1-r       & \quad t<(1-r) \\
  t  & \quad (1-r)<t<(1+r) \\
  1+r & \quad t>(1+r).
  \end{cases} 
   \]
Integrating the above equation, we have 
 \begin{equation} \label{eq3}
 \frac{dQ}{dt}=
  \frac{2\pi\kappa Q}{r}
  \begin{cases}
  0       & \quad t<(1-r) \\
  \ln{\frac{t}{1-r}}  & \quad (1-r)<t<(1+r) \\
   \ln{\frac{1+r}{1-r}} & \quad t>(1+r).
  \end{cases}
  \end{equation}
If we ignore compression and assume heat capacity of the matter, $ C_v = $ constant, we can write
\begin{equation*} \label{eq4} 
dT=\frac{dQ}{C_v} \Rightarrow T(r,t)=\frac{1}{C_v}\int \Big( \frac{dQ}{dt}\Big)dt
\end{equation*}
$$ T(r,t)=\frac{2\pi\kappa Q}{C_{v} r} \int_{0}^{t} dt
 \begin{cases}
 0       & \quad t<(1-r) \\
 \ln{\frac{t}{1-r}}  & \quad (1-r)<t<(1+r) \\
 \ln{\frac{1+r}{1-r}} & \quad t>(1+r).
 \end{cases}  $$\\
On solving the above integral, we finally have 
 \begin{equation} \label{eq7}
 T(r,t) =\frac{2\pi\kappa Q}{C_{v} r}
 \begin{cases}
 0      &\quad t< (1-r)\\
 t\ln\frac{t}{1-r}-t+1-r & (1-r)<t<(1+r)\\
 t\ln\frac{1+r}{1-r}-2r & \quad t>(1+r).
 \end{cases}
\end{equation}
Thus if $ t>1+r $ and $r\rightarrow 0 $ then
 $$ T(0,t)=\lim_{r \to 0} T(r,t),$$
which can be simplified as 
$$ T(0,t) =\frac{2\pi\kappa Q}{C_{v}}\lim_{r \to 0} \Bigg[t\frac{\ln{(1+r)}-\ln{(1-r)}}{r}-2\Bigg]$$
\begin{equation}\label{eq8} 
\Rightarrow T(0,t)=\frac{4\pi\kappa Q}{C_{v}}(t-1).
    \end{equation}
The discontinuity surface is determined by contour $ T(r,t)=T_{c}$. 
The critical point $(r_{c},t_{c})$, at which the SL and TL discontinuities converge, is determined by the condition
\begin{equation*} 
\Bigg(\frac{\partial r}{\partial t}\Bigg)_{T_{c}}  =  \Bigg(\frac{\partial T}{\partial t}\Bigg)_{T_{c}}\Bigg/\Bigg(\frac{\partial T}{\partial r}\Bigg)_{T_{c}}=1.
\end{equation*}
Substituting the value for $t > (1+r)$ from eqn \ref{eq7} we have
\begin{gather}\label{eq11}
 t_{c}=\frac{\ln{\frac{1+r_{c}}{1-r_{c}}}}{\Bigg(\frac{2}{1-r_{c}^{2}}-\dfrac{1}{r_{c}}\ln{\dfrac{1+r_{c}}{1-r_{c}}}\Bigg)}, \\ \nonumber
 =\Bigg\{\Bigg[(1-r_{c}^{2})\ln{\bigg(\frac{1+r_{c}}{1-r_{c}}\Bigg)^{\frac{1}{2}}}\Bigg]^{-1}-\frac{1}{r_{c}}\Bigg\}^{-1}.
\end{gather}

\section{TL detonation in the presence of magnetic field}

To add a magnetic field in the first law of Thermodynamics, we use Maxwell's fields.
The energy generated within a volume V in time $\delta t$, by an electric field $\varepsilon $ acting on  current density $\jmath$ is given by \citep{wasserman}
$$\delta W=-\delta t\int_V \jmath\cdot\varepsilon\;\mathrm{d V} $$
For the quasi static and reversible system, work done by the system
 \begin{equation}\label{eq12}
 \delta W = \delta t \int_V \jmath\cdot\varepsilon\;\mathrm{d V}.
 \end{equation}
 Using Maxwell's Equation (Ampere's law  in differential form) and after simplification, we get
 \begin{equation}\label{eq15}
 \begin{split}
   \delta W =\delta t\bigg[\frac{c}{4\pi}\bigg\{\int \,\Delta\cdot(H\times\varepsilon)\;dV \\
 +\int H\cdot(\Delta\times\varepsilon)\,dV\bigg\} 
 -\frac{1}{4\pi}\int \frac{\delta D}{\delta t}\varepsilon\,dV\bigg].
 \end{split}
 \end{equation}
 The second term of the RHS 
 $\int \Delta\cdot(H\times\varepsilon)\,dV$ can be written in terms of the surface integral using Gauss Theorem. For large distances the surface integral can be neglected.
 
 \begin{figure} 
\includegraphics[width = 1.75in,height=1.7in]{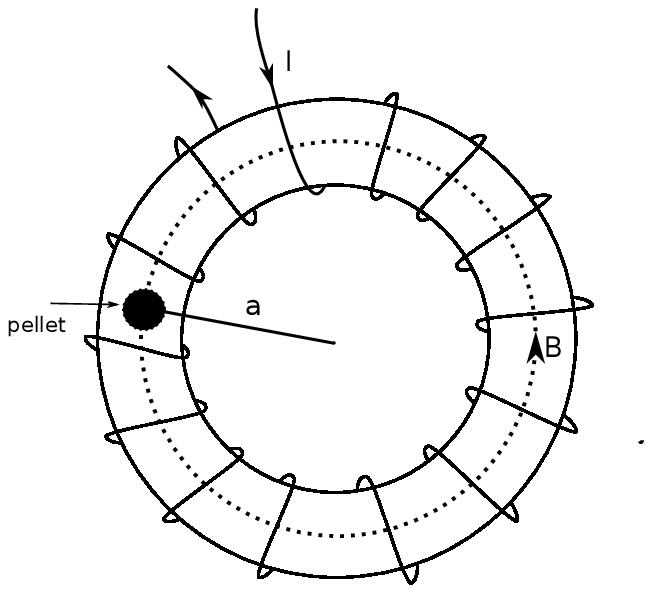}
\includegraphics[width = 1.25in,height=1.75in]{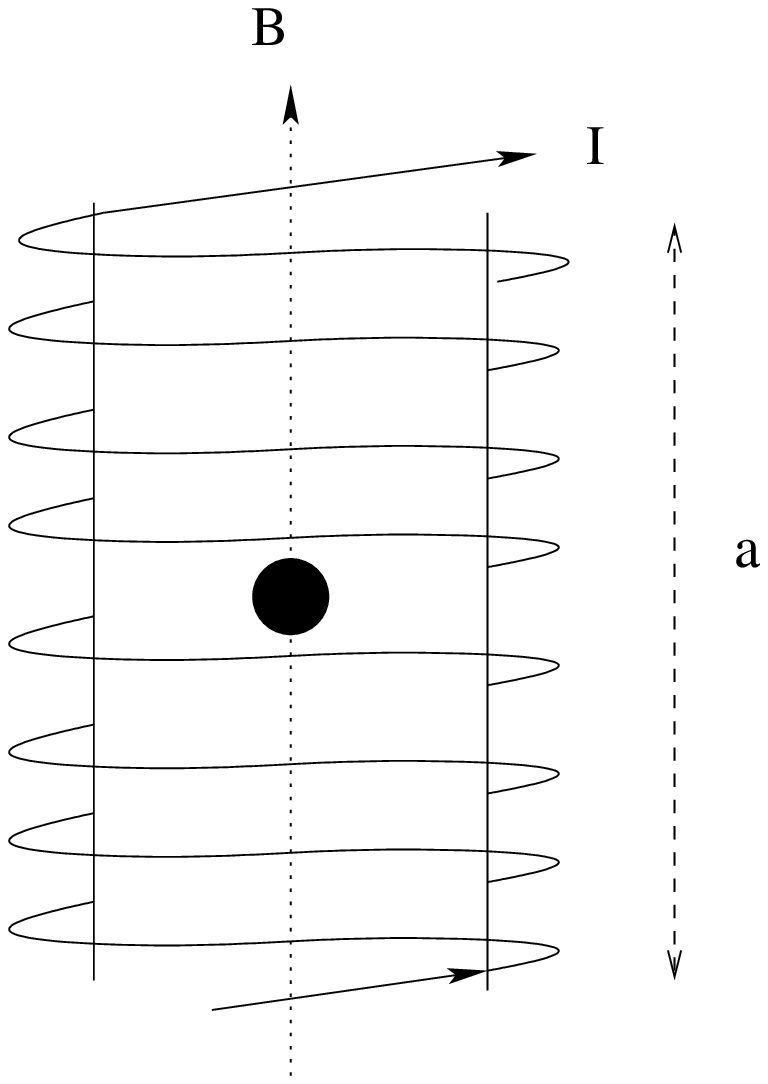}\\
(a)  \hspace{3cm} (b)
\caption{Figure describing the (a) Toroid and (b) Solenoid coils giving rise to a spatially constant magnetic field along the pellet. 
The magnetic field is constant as $a$ is assumed to be much greater than $R$. The magnetic field
inside the pellet is therefore uni-directional and constant. The current $I$ is time-varying.}
\label{toroid}
\end{figure}
 
Using Maxwell's Equation (Faraday's law) and taking only magnetic field part, eqn \ref{eq15} reduces to
\begin{equation}\label{eq16}
\delta W = -\dfrac{1}{4\pi}\int_V H\cdot\delta B\, dV
\end{equation}
Using the definition of magnetization density $\mathcal{M}$, as $H=B-4\pi \mathcal{M}$, eqn \ref{eq16} takes the form
$$ \delta W = -\dfrac{1}{4\pi}\bigg[\int_V B\,\delta B\;dV+4\pi \int_{V^{'}}  \mathcal{M}\,\delta B\;dV\bigg]. $$ 
Here, the term $-\dfrac{1}{4\pi}\int_V B\,\delta B\;dV$ is total field energy integrated over all space 
and can be absorbed into the internal energy \citep{wasserman}. The second term $ \int_{V^{'}}  \mathcal{M}\,\delta B\;dV$ is the integral over 
the volume of magnetized matter and the term of our interest. Talking only about the work done by the magnetized matter and defining
\begin{equation}
 \int_{V^{'}}  \mathcal{M} dV = M,
\end{equation}
the total work done by the magnetic field is given by
\begin{equation}\label{eq17}
\bar{d} W= -M\,dB.
\end{equation}
The first law of Thermodynamics  can be written as (ignoring the compression)
\begin{equation}\label{eq18}
d Q=M dB.
\end{equation}

\begin{figure} 
\includegraphics[width = 2.5in]{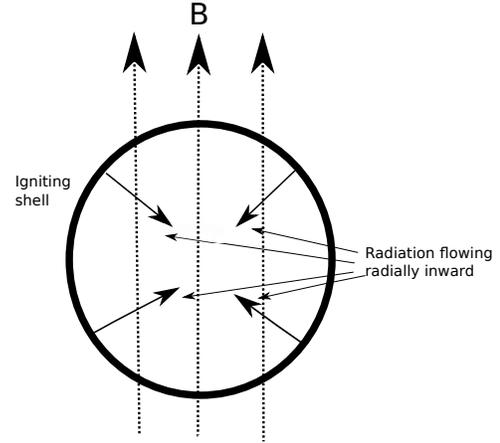}
\caption{The radiation and the magnetic field direction are shown inside the pellet. The magnetic field is uni-directional whereas, the radiation is symmetric.}
\label{fig-prof}
\end{figure}

Therefore, if our system has both radiation and magnetic field effect, the total heat transfer in the system given by two process, one due to radiation and other 
due to magnetic field, and is given by
$$ dQ_{Total}=dQ_{Radiation}+ dQ_{magnetic field} $$
 \begin{equation}\label{eq19}
 dQ_{total}=\bigg(\frac{dQ}{dt}\bigg)dt+M dB.
 \end{equation}
 
Assuming that the matter interacts with the magnetic field, the average magnetization can be given by
$$M   = \frac{N\mu^{2} B}{k_{B} T}$$. We should also mention that as we have neglected the compression of the fuel, we do not have any multiplicative term involving radiation and magnetic field. This is to say that we have neglected the interaction between the two processes.
Substituting M in eqn \ref{eq19}, we have
\begin{widetext}
\begin{equation}\label{eq20}
 dQ_{total}= \frac{2\pi\kappa Q}{r}
\begin{cases}
0       & \quad t<(1-r) \\
\ln{\frac{t}{1-r}}  & \quad (1-r)<t<(1+r) \\
\ln{\frac{1+r}{1-r}} & \quad t>(1+r).
\end{cases}
\Bigg\}\,dt+\frac{N\mu^{2} B}{k_{B} T}dB
\end{equation}
and the temperature profile can be obtained from the equation
\begin{equation}
 C_vdT=dQ_{total}.
 \label{cvdt}
\end{equation}
Inserting the value of $dQ_{total}$ from eqn \ref{eq20} in eqn \ref{cvdt}, we have 
\begin{equation}\label{eq21}
 dT= \frac{2\pi\kappa Q}{rC_v}
\begin{cases}
0       & \quad t<(1-r) \\
\ln{\frac{t}{1-r}}  & \quad (1-r)<t<(1+r) \\
\ln{\frac{1+r}{1-r}} & \quad t>(1+r).
\end{cases}
\Bigg\}dt+\frac{N\mu^{2} B}{C_v k_{B} T}dB
\end{equation}
\end{widetext}
The above equation implies that only a time varying magnetic field can contribute to the heat (or the temperature). Therefore, we discuss our results for a 
sinusoidal varying magnetic field.

Figure \ref{toroid} shows the schematic of a model experimental setup, illustrating the pellet (core surrounded by the shell) in an external magnetic field. 
In (a) the pellet is kept inside a toroid, with the radius of the toroid $a$ being sufficiently larger than the dimension of the pellet and in (b) the pellet is placed inside a long solenoid (length $a$ of the solenoid being $>> R$). Then the magnetic field $B$ inside the pellet can be assumed to be spatially constant. The close up of the pellet is illustrated in fig. \ref{fig-prof}, showing the cross-section of the pellet. The spatially constant magnetic field direction
and $4 \pi$ radiation inside the pellet is explained in detail.
 
The magnetic field is generated by a toroid or a solenoid. The direction and the magnitude of the magnetic field is approximately 
constant inside the pellet (core surrounded by the shell) and is defined as
\begin{equation}\label{eq22}
  B=B_0 \sin{(\omega\, t)} \Rightarrow \quad dB=B_0\,\omega\cos{(\omega\, t)} dt 
  \end{equation}
In the case of toroidal magnetic field, $B_0=\frac{\mu_{0} NI_0}{2\pi a}$ , where $a$ is the distance of the pellet from the center of the toroid, $\mu_0$ is 
the permeability in free space, $I_0$ is the current and $N$ is the total number of turns. It can also be written as $B_0=\mu_0nI_0$, with $n$ being the turn density
defined as $n=N/2 \pi a$. The solenoid magnetic field can also be written as, $B_0=\mu_0nI_0$ ,where $n$ number of turns per unit length defined as $n=N/a$. 
However, here the $a$ is defined as the length of the solenoid.
The magnetic field and the position of the pellet are shown in the figures \ref{toroid} and \ref{fig-prof}.
Using equation \ref{eq21} and \ref{eq22}, we find differential equation for temperature profile inside the pellet
  \begin{widetext}
  \begin{equation}
   \frac{dT(r,t)}{dt}= \frac{K_1}{r}
  \begin{cases}
  0       & \quad t<(1-r) \\
  \ln{\frac{t}{1-r}}  & \quad (1-r)<t<(1+r) \\
  \ln{\frac{1+r}{1-r}} & \quad t>(1+r).
  \end{cases}
  \Bigg\}+\frac{K_2\omega}{T(r,t)}\sin(\omega\,t)\cos(\omega\,t)
  \label{sin-eq}
  \end{equation}
 \end{widetext}
 where $$ K_1=\frac{2\pi\kappa Q}{C_v}\quad \text{and}\quad K_2=\frac{N\mu^{2} B_0^2}{C_v k_{B}}$$.
 In our calculation, we have taken the temperature $T$ in units of $K_1$, and for convenience, it is assumed to be $1$.

If the magnetic field is varying very slowly such that we can approximate $\sin(\omega\,t)\approx\omega\,t$, then equation \ref{sin-eq} takes the form,
 \begin{widetext}
     \begin{equation}
    \frac{dT(r,t)}{dt}= \frac{K_1}{r}
     \begin{cases}
     0       & \quad t<(1-r) \\
     \ln{\frac{t}{1-r}}  & \quad (1-r)<t<(1+r) \\
     \ln{\frac{1+r}{1-r}} & \quad t>(1+r).
     \end{cases}
     \Bigg\}+K
     _2\frac{\omega^2\,t}{T(r,t)}.
     \label{slow-eq}
     \end{equation}
     
 \end{widetext}
Both the above two equations are non-linear and can be solved numerically.

\section{Results}

In our calculation, we have neglected the compression of the core. As compression is not taken into account, we also have neglected any interaction 
between the radiation and magnetic field.
Also, the effect of the magnetic field comes through the work done by the magnetic field. The heat generated by the fusion of the pellet and that from the magnetic field is added, and then from the equation of state (EoS), the final temperature is calculated. In previous studies and experiments, such effect of the magnetic field has not been studied. In those studies, the effect of the magnetic field comes from the compression of the fuel.

Recently, in the Omega facility \citep{chang, hohenburger}, axial magnetic field was applied to a laser-driven ICF, where the heat losses were suppressed considerably and thereby heating of the ion temperature to higher values. In MagLIF \citep{slutz, cuneo, gomez}, the cylindrical implosion with axial magnetic field was carried out. The axial magnetic field reduces the thermal loss throughout the implosion region. The laser first heat the fuel and the axial magnetic field compresses and heat the fuel further to a 
higher temperature.
The magnetic field helps in confining and compressing the fuel to higher temperatures. However, the work done there is only $PdV$. In our case, the work is both from $PdV$ and $MdB$.
As we have neglected the compression of the fuel to keep our calculation simple, the temperature does not rise due to the compression of the fuel. However, it should be mentioned that in actual 
experiments both effects had to be taken into account to calculate the total heat and the final temperature.

The basic idea of this present work is that as the shock wave propagates inward, the shock shifts from SL to TL shock. Once this happens, the likelihood of instabilities decreases and thus the effectiveness of ICF increases. It is, therefore, necessary that we calculate the point of this transition from SL to TL shocks. 
Previously \citep{csernai,csernai2}, the point of SL to TL shock has been calculated where the radiation effect is only taken into account. The solution of the problem could
be obtained analytically (solving eqn. \ref{eq7} and eqn. \ref{eq11}). However, when we take both the radiation and the magnetic effect the problem cannot be solved analytically 
anymore. Therefore, we solve
eqn. \ref{sin-eq} and eqn \ref{slow-eq} numerically. 

\begin{figure} 
\includegraphics[width = 3.5in]{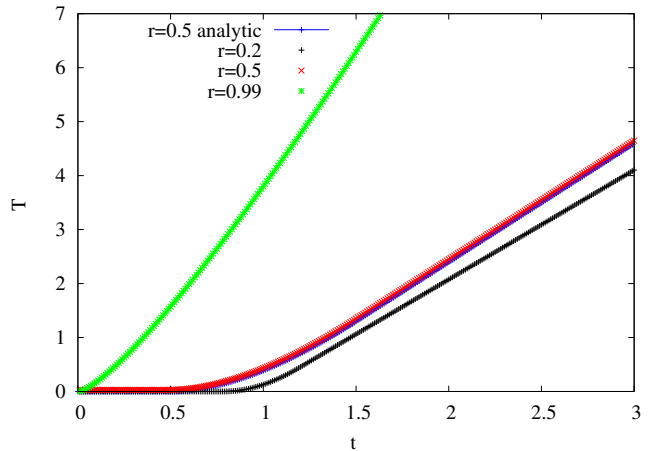}
\caption{The variation of temperature $T$ with time $t$ is shown. Curves are shown for three different radial distance $r$, where $r=0.2, 0.5$ and $0.99$. The black-$+$ curve is 
for $r=0.2$, the red-$\times$ curve is for $r=0.5$ and the green-$*$ curve is for $r=0.99$. The magnetic field is zero for this figure. The analytically solved curved 
is marked in blue-$|$ and for $r=0.5$. }
\label{no-mag}
\end{figure}

First, we have solved eqn. \ref{sin-eq} numerically to obtain temperature $T$ as a function of time $t$ for some given $r$. To match and verify that our numerical procedure is satisfactory we have first solved the equation for zero magnetic fields. The numerical solution of eqn. \ref{sin-eq} (for $B=0$) should match with the analytical curve. 
The numerical solution of eqn. \ref{sin-eq} is shown in fig \ref{no-mag}. We have plotted the $T-t$ curve for three different values of $r$, $r$ being $0.2, 0.5$ and $0.99$. 
The curve for $r=0.5$ obtained numerically matches well with the analytically obtained curve for same $r$.
Initially, for some time curve has zero temperature, then after $t>(1-r)$ there is a logarithmic rise in the 
curve (consistent
with eqn. \ref{sin-eq}), and then the curve becomes almost a straight line. For $r=0.2$, $1-r$ takes a large value $0.8$, and so the zero temperature extends till $t=0.8$ and 
the logarithmic part is only between $0.8 - 1.2$. For higher values of $t$ the curve becomes a straight line. For $r=0.5$ the $T=0$ region is up-to $t=0.5$ and the logarithmic part extends from 
$t=0.5$ to $1.5$ and then the curves becomes a straight line. For $r=0.99$ there is almost no $T=0$ curve, the logarithmic curve extends from $t=0 - 1.99$ and then it is a 
straight line.
The slope of the curves depends on the extent of the zero temperature and logarithmic part, and the curve with the largest zero temperature region has the least slope and vice-versa.

\begin{figure} 
\includegraphics[width = 3.5in]{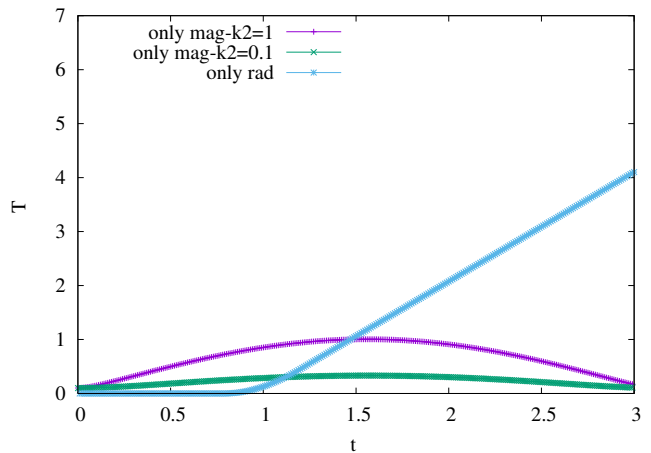}
\caption{The variation of temperature $T$ as a function of time $t$ is shown. Only the temperature due to the magnetic field is shown in the curves marked as only-mag 
(purple-slash and green cross). The graph is plotted for different magnetic field strength, the smaller being one-tenth of the larger. The field strength is given in terms 
of the constant $K_2$. For comparison with the radiative contribution, the radiation temperature curve is also plotted and marked as only-rad (blue-star). The radiation curve is plotted for $r=0.2$.}
\label{mag-only}
\end{figure}

The above results show only the radiation contribution to the heat and thereby to the temperature. The magnetic field (without any radiation effect is shown in fig \ref{mag-only}.
For comparison with the radiation contribution, we have also plotted the radiation plot for $r=0.2$. Although, we have assumed a sinusoidal magnetic field the frequency of variation is not very high. The curve marked with only mag-$K_2=1$ is the curve obtained with $K_2=K_1=1$. 
Similar to the sinusoidal nature the heat due to magnetic field variation first rises with time till $t= \pi/2$. 
Then the heat falls off with time and reaches zero at $t=\pi$.
The negative part of the sinusoidal variation 
would make the temperature imaginary, and therefore, the $T$ for the negative part is assumed to be zero.
However, for our analysis, the curves are shown up to $t=3$, $< \pi$ which is sufficient for our calculation as the contour plots saturate before $t=3$ (can be seen later).

The temperature curve also follows such behaviour as clear from fig \ref{mag-only}. The fluctuating magnetic field leads to a 
fluctuating temperature however the fluctuation is small. Initially, as the magnetic field rises with time the heat due to the magnetic field also grows, and that is quite natural. 
However, after $t=\pi/2$, as the magnetic field decreases the heat also decreases, that is the contribution due to the magnetic field goes down. However, it is expected that the heat will need some time to equilibrate and will not directly go down with a magnetic field. For an ideal case, our assumption is valid for the dynamical process at the initial time and not beyond times after thermalization had taken place. However, that will not affect our result a lot because from the figure it is clear that after $t=\pi/2$, the radiation heat will dominate the total heat or temperature of the process. 
The radiation curve is plotted for $r=0.2$, and for larger $r$ the effect due to magnetic field in $T$ beyond $t=\pi/2$ will reduce further.

The strength of the magnetic field can be controlled by choosing the value of $K_2$. We have decreased the magnetic field strength by choosing $K_2=0.1$ (one-tenth 
of the previous value). With such a choice of $K_2, $ the effect due to the fluctuating magnetic field reduce even further,
and it becomes almost a straight line in comparison to the radiation temperature. Therefore, the fluctuating magnetic field will not change our result a great deal.

\begin{figure} 
\includegraphics[width = 3.5in]{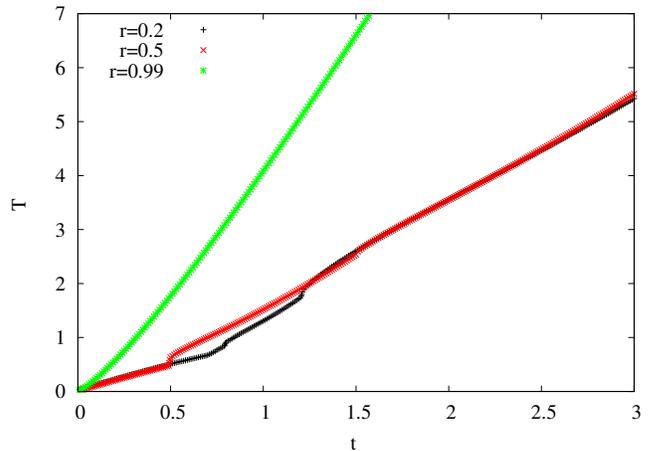}
\caption{The temperature $T$ variation with time $t$ is plotted in the figure. Curves are shown for three different radial distance $r$, where $r=0.2, 0.5$ and $0.99$, with the marking of the 
curves remaining the same as of fig \ref{no-mag}. The constant $K_2$ is of the same value as of that for the radiation $K_1=1$, which means that both effect have almost equal 
contribution.} 
\label{sin}
\end{figure}

Once, it is established the numerical solving procedure is quite good, we then move on to solving eqn. \ref{sin-eq} having a contribution from radiation and magnetic field.
First, we choose that the heat contribution from the magnetic field is assumed to be of the same order as that of the radiation ($K_1=K_2=1$). 
In fig \ref{sin} we plot the $T-t$ curve for three different values of $r$ as done for the nonmagnetic case. For, $r=0.2$ the $T$ at small $t$ (from $0$ to $0.8$) 
is not zero as there is some contribution from the magnetic field. The contribution from magnetic field is continuous but the contribution from the radiation is discontinuous 
and it is discontinuous at two points ($t < 1-r$ and $t> 1+r$). Therefore the final curve has some discontinuous jumps at these two points. The disconnected jumps 
are most prominent in $r=0.2$ curve. The jump or the discontinuity is also there in the $r=0.5$ curve but for $r=0.99$ curve the jump is almost absent as the condition $t < 1-r$
for the radiation is almost nonexistent.

\begin{figure} 
\includegraphics[width = 3.5in]{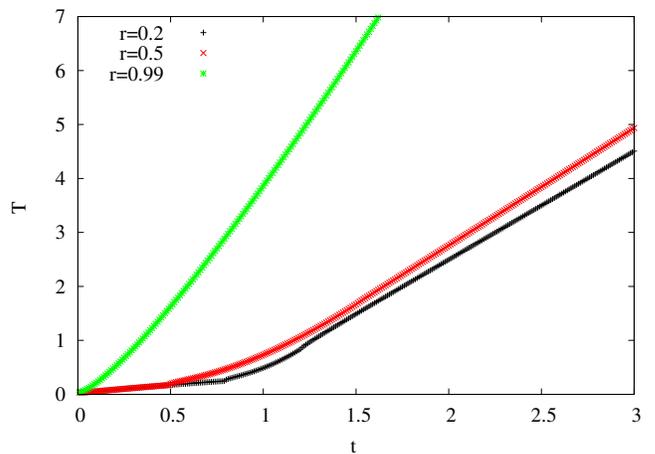}
\caption{The variation of temperature $T$ with time $t$ is shown in the figure. Curves are shown for three different radial distance $r$, where $r=0.2, 0.5$ and $0.99$, with the marking of the 
curves remaining the same. The constant $K_2$ is taken to be $0.1$ which means that the radiation effect dominates over the magnetic effect.} 
\label{sin-1o}
\end{figure}

In fig \ref{sin-1o} we have plotted the same curve with the assumption that the constant term of the magnetic contribution is $0.1$ times that of the previous case ($K_2=0.1$). This is to check what the effect of the strength of the magnetic field has on the conversion of SL to TL curve. The plot of $T$ vs. $t$ for such a case is close to that of the nonmagnetic case, 
which is also consistent. The discontinuity of the curve in the different regions are less prominent than the previous one, and the curve is quite smooth.

\begin{figure} 
\includegraphics[width = 3.5in]{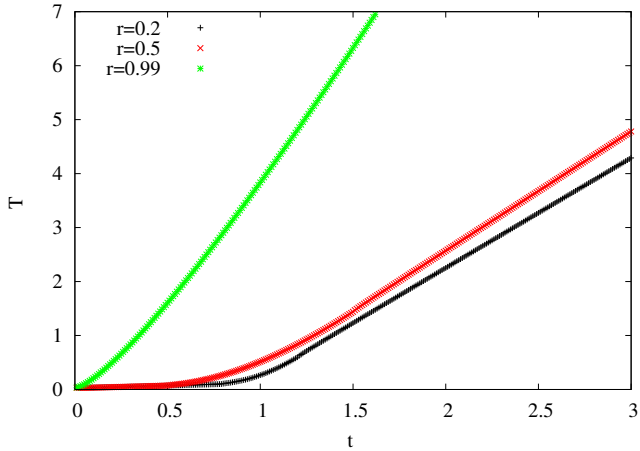}
\caption{The temperature $T$ vs. time $t$ is illustrated. Curves are shown for three different radial distance $r$, where $r=0.2, 0.5$ and $0.99$, with the marking of the 
curves remaining same. The curve is obtained by solving eqn. \ref{slow-eq}, i.e. for small $\omega$ approximation.} 
\label{small}
\end{figure}

Finally, the last $T-t$ curve (fig \ref{small}) is plotted with the assumption that the magnetic field is not varying fast and the $\omega$ is small. Solving eqn. \ref{slow-eq} for
the $\sin(\omega\,t)\approx\omega\,t$, 
with $\omega$ being treated as a constant. The curve almost replicates the nonmagnetic curve which is also what is expected. 

The magnetic field profiles that we have assumed can be implemented using the configuration of a toroid or a solenoid. The toroidal shape is difficult to be realized in actual 
experimental setup.
However, the solenoidal configuration can be realized similar to the set up of that of MagLIF. There, instead of a conductor of a cylindrical shape, we can have a solenoid.
The sinusoidal magnetic field can be obtained by a sinusoidal current. In the case of the solenoidal setup, we can even get rid of the compression if the current density
is parallel to the magnetic field. However, to increase the efficiency of the ICF, we want the compression to happen along with the magnetic work. Therefore, for that case, we want to choose some other configuration in the actual experimental set-up.

\begin{figure} 
\includegraphics[width = 3.5in]{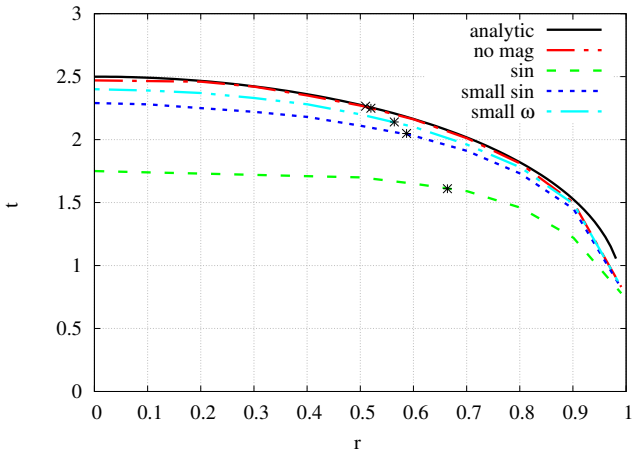}
\caption{$t$ as a function of $r$ are shown in the figure. The star marked on the curves denotes the point of conversion of SL to TL shocks, that is to say $T(r,t)=T_{c}$.
The curve marked analytic denotes the curve obtained from the analytical solution, the red-dash-dot curve is the numerical solution for the nonmagnetic case. The three magnetic 
curves are drawn as follows: the sinusoidal magnetic field with $K_2=1$ is marked with sin (green-dash), the curve with smaller field strength ($K_2=0.1$) is marked as 
small-sin (blue-dot) and finally the slowly varying approximation curve with the $\omega$ constant is marked as small-omg (cyan-dash-double-dot). All the curves are plotted for $T=3$.}
\label{t-r}
\end{figure}

After obtaining the $T-t$ curve, we can then plot the $t-r$ curve for all of the above cases. This can be done by plotting $t$ as a function of $r$ for some fixed temperature. 
In fig \ref{t-r} we have plotted $t$ as a function of $r$ for $T=3$. As evident from the figure, the analytic and the numeric curve (for $B=0$) almost overlap with each other, assuring that the numerical procedure is quite good. The point of transition from SL to TL shocks is calculated from the fact that at that point the slope of the curve should be $\pm 1$.
The star marked on the curves signify those transition points. 
The point ($r_c,t_{c}$) separates the SL and TL part of the discontinuity surface $T(r,t)=T_{c}$. 

Initially a shock is formed at $r=R $ at time $t=0$ and then 
propagates inward. 
The process initially proceeds slowly, but then accelerates up by the radiative heat transfer and at $r_{c}$ it goes over smoothly into a TL discontinuity.
The region inside the marked star point is the TL region and the region outside is the SL region. 
For the non-magnetic implosion, for $ r_{c}=0.51 $ the critical time comes out to be $t_c=2.26$ (from eqn \ref{eq11}) and the critical temperature $T(r_{c},t_{c})$ is 
 $$ T(r_{c},t_{c})=\frac{2\pi CQ}{C_{v} r_{c}}\Bigg[t_{c}\ln{\frac{1+r_{c}}{1-r_{c}}}-2 r_{c}\Bigg] $$
 $$=3\bigg(\frac{2\pi CQ}{C_{v}}\bigg)=3 $$
(since $K_1=\frac{2\pi CQ}{C_v}=1$).
The time required to heat the center of core up to a temperature $T_{c}$ (from eqn \ref{eq8}), is
$$ T_{c}=\frac{4\pi CQ}{C_{v}}(t-1)\Rightarrow 2(t-1)=3 $$
$$ \Rightarrow t=2.5 $$
Numerical analysis for $T=T_C=3$ gives the critical point to be $r_c=0.52$, $t_c=2.25$, which matches quite well with analytic values.

For the magnetic field induced implosion, we had three cases. The first being the sinusoidal magnetic field having $K_2=K_1=1$, second also for a sinusoidal field but 
with field strength smaller than the first case ($K_2=0.1$), and the third calculation is done with small $\omega$ approximation.
The sinusoidal curve with $K_2=1$
lies much below the nonmagnetic curve which signifies that the time taken to reach a temperature of $T=3$ is much less as compared to the nonmagnetic case. 
This is also quite natural 
as the heat from both radiation and the magnetic field contribute, and for a particular point to reach the desired temperature requires smaller time. 
The green-dash curve is plotted for a sinusoidal field with $K_2\approx K_1=1$ and blue-dot curve with sinusoidal field having $\frac{K_2}{K_1}\approx 0.1$. 
Using the slope of contour we find the critical point for both the curve : $r_c=0.664$, $t_c=1.6364$ 
for the green-dash curve and $r_c=0.587$,$t_c=2.091$ for blue-dot curve. 
The slowly varying magnetic field approximation case is plotted with the cyan-dash-double-dot curve. 
In this case critical point comes out to be $r_c=0.56$ and $t_c=2.15$. 

For the sinusoidal curve having a significant contribution to the TL shock region of the core extends further out. 
As the TL region extends further the chances of instabilities appearing in the region reduces, and the heating becomes more efficient.
The curve with smaller magnetic field strength lies above the curve with higher field strength. The time taken to heat a 
particular radial point is longer than the previous case. As the efficiency of the heating due to the magnetic field decreases with a decrease in magnetic field strength, 
the time taken is longer. 
The transition point ($t_c,r_c$) also shifts inward to the core. That is to say that a smaller portion of the core now has TL shocks.
This feature is further evident when we plot the curve for slowly varying magnetic field approximation. The magnetic field varies linearly with time, but the strength of the field is much
smaller.
The curve shifts to higher $t$ values and lies close to the nonmagnetic curve. The efficiency of the magnetic heating decreases further, and most of the contribution 
in the heating comes from the radiation.
The transition point shifts much inward (the $r_c$ is slightly larger than the nonmagnetic case).

Therefore, from fig \ref{t-r} we can conclude that the heating due to the magnetic field can be quite significant in determining the transition point of SL to TL shocks and can be exploited 
to minimize the instabilities that reduce the efficiency of the ICF.

\section{Summary \& Conclusion}
To summarize, we have studied the effect of time-varying magnetic field on the smooth and continuous transition from SL to TL detonation in an implosion induced by radiation.
The heating of the core arises due to the rapidly igniting explosive shells which surround it. The central region is swiftly heated to a very high temperature by incoming radiation from all directions. On top of this, we employ a time-varying magnetic field at the core. A constant magnetic field doesn't affect the heating dynamics of the core.

The magnetic field has a significant effect on the dynamics of continuous transition from SL to TL detonation depending on the strength of the magnetic field.
In this analysis, we have employed a sinusoidally varying magnetic field which can be realized in laboratories. We have neglected the compression of the core due to the magnetic field. 
The effect of the magnetic field comes through the work done by the magnetic field ($MdB$). The heat generated by the radiation (which can be obtained by the fusion of the pellet or by controlled laser irradiation) and that from the magnetic field is added, and then the temperature is calculated. In previous studies and experiments carried out in laboratories such effect due to magnetic work has not been taken into account. In those studies, the effect of the magnetic field comes from the compression of the fuel.
There the magnetic field helps in confining and compressing the fuel to rise to a further higher temperature. In our case, the work is both from radiation ($PdV$) and magnetic ($MdB$).
We have neglected the compression of the fuel to keep our calculation simple which could be realized even in the experimental set up if the magnetic field and the current density are parallel. However, to increase the efficiency of the ICF, we want the compression to happen along with the magnetic work. Therefore, for that case, some other configuration needs to be sought. The MagLIF setup can be used to have all such effects.

The basic idea of this present work is that as the shock wave propagates inwards, there is a transition from SL to TL shocks. Once this happens, the likelihood of instabilities 
decreases further and thus the effectiveness of ICF increases. The change in the core volume of TL detonation gets rid of unwanted RT instabilities.
The effect of the magnetic field can have enormous significance 
particularly in the experimental
\citep{casey, hora, hurricane, park} and theoretical \citep{kasotakis,atenzi,csernai-new} study of ICF of fuel. 
However, we should mention here that we have assumed a model set up for the generation of the toroidal or solenoidal magnetic field. In the actual experiments, 
more complex set up may be needed for the fact that the pellet is to be heated by lasers from all $4 \pi$ directions along with magnetic field generation. 
Also, the effect of the induced electric field and radiation is not taken into account.

We should also mention that we have assumed a fluctuating magnetic field and that expresses itself in the form of fluctuating heat or fluctuating temperature. 
However, to be precise, this assumption is valid till thermalization takes place, that is in the initial phase of this dynamical process.
However, in our calculation, the value of the frequency we choose makes the fluctuation is minimal, and after some time the radiation effect dominates over the magnetic effect. 
The fluctuating magnetic field would, therefore, not change our result a great deal. 
The magnetic field effect can also affect the hadronization of QGP volume involving detonation in high energy experiments. The emergence of a discontinuity due to the 
application of an oscillating magnetic field is a phenomenon which has not been observed or expected, and further study is needed to understand it, which is our present endeavor.

\acknowledgments
RM is grateful to the SERB, Govt. of India for monetary support in the form of Ramanujan Fellowship (SB/S2/RJN-061/2015) and Early Career Research Award (ECR/2016/000161). 
RM and SS would also like to thank IISER Bhopal for providing all the research and infrastructure facilities.


\section*{References}

\begin{thebibliography}{}

\bibitem[\protect\citeauthoryear{Atenzi et al.} {2014}]{atenzi}{\bf Atenzi, S.,  Ribeyre, X.,  Schurtz, G.,  Schmitt, A. J.,  Canaud, B.,  Betti, R. and  Perkins, L. J.} (2014). Shock ignition of thermonuclear fuel: principles and modelling, Nucl. Fusion 54, 054008 \\
\bibitem[\protect\citeauthoryear{Casey et al.} {2014}]{casey} {\bf Casey, D.T.,  Smalyuk, V. A.,  Raman, K. S.,  Peterson, J. L.,  Berzak Hopkins, L., Callahan, D. A.,  Clark, D. S.,  Dewald, E. L.,  Dittrich, T. R.,  Haan, S. W.,  Hinkel, D. E.,  Hoover, D., Hurricane, O. A., Kroll, J. J.,  Landen, O. L.,  Moore, A. S., Nikroo, A., Park, H.-S., Remington, B. A., Robey, H. F.,  Rygg, J. R., Salmonson, J. D.,  Tommasini, R., and  Widmann, K.} (2014). Reduced instability growth
with high-adiabat high-foot implosions at the National Ignition Facility, Phys. Rev. E 90, 011102(R) \\
\bibitem[\protect\citeauthoryear{Chang et al.} {2011}]{chang} {\bf Chang, P. Y., Fiksel,  G., Hohenberger, M.,  Knauer, J. P., Betti, R. , Marshall, F. J.,  Meyerhofer, D. D.,  Séguin, F. H., and  Petrasso, R. D.} (2011). Fusion Yield Enhancement in Magnetized Laser-Driven Implosions, Phys. Rev. Lett. 107, 035006 \\
\bibitem[\protect\citeauthoryear{Csernai} {1987}]{csernai} {\bf Csernai, L.P.} (1987). Detonation on a timelike front for relativistic systems, Zh. Eksp. Teor. Fiz. 92, 379-386. Sov. JETP 65, 216-220 \\
\bibitem[\protect\citeauthoryear{Csernai} {1994}]{csernai2}{\bf Csernai, L.P.} (1994). Introduction to Relativistic Heavy Ion Collisions. (Chichester: Wiley)\\ 
\bibitem[\protect\citeauthoryear{Csernai et al.} {2015}]{csernai3} {\bf Csernai, L. P. and Strottmann, D. D.} (2015). Volume ignition via time-like detonation in pellet fusion, Laser \& particle beam 33, 279 \\
\bibitem[\protect\citeauthoryear{Csernai et al.} {2018}]{csernai-new} {\bf Csernai, L.P., Kroo, N.\&  Papp , I.} (2018).
Radiation dominated implosion with nano-plasmonics,  Laser and Particle Beams 1-8 \\
\bibitem[\protect\citeauthoryear{Cuneo et al.} {2012}]{cuneo} {\bf Cuneo, M. E.,  Herrmann, M. C., Sinars, D. B., Slutz, S. A., Stygar, W. A., Vesey, R. A.,
Sefkow, A. B., Rochau, G. A., Chandler, G. A., Bailey, J. E., Porter, J. L., McBride, R. D., Rovang, D. C.,
Mazarakis, M. G., Yu, E. P.,  Lamppa, D. C., Peterson, K. J.,  Nakhleh, C., Hansen, S. B., Lopez, A. J., Savage, M. E.,
Jennings, C. A., Martin, M. R., Lemke, R. W., Atherton, B. W., Smith, I. C., Rambo, P. K.,  Jones, M., Lopez, M. R.,
Christenson, P. J., Sweeney, M. A.,  Jones, B., McPherson, L. A.,  Harding, E., Gomez, M. R., Knapp, P. F., Awe, T. J.,
Leeper, R. J., Ruiz, C. L., Cooper, G. W., Hahn, K. D., McKenney, J., Owen, A. C., McKee, G. R., Leifeste, G. T.,
Ampleford, D. J., Waisman, E. M., Harvey-Thompson, A., Kaye, R. J., Hess, M. H.,  Rosenthal, S. E., and  Matzen, M. K.} (2012). Magnetically Driven Implosions for Inertial
Confinement Fusion at Sandia National Laboratories, IEEE Trans. Plasma Sci. 40, 3222\\
\bibitem[\protect\citeauthoryear{Gomez et al.} {2014}]{gomez} {\bf Gomez M.R., Slutz, S. A., Sefkow, A. B., Sinars, D. B., Hahn, K. D.,  Hansen, S. B., Harding, E. C.,  Knapp, P. F., Schmit, P. F., Jennings, C. A., Awe, T. J.,  Geissel, M.,  Rovang, D. C., Chandler, G. A., Cooper, G. W.,  Cuneo, M. E., Harvey-Thompson, A. J.,  Herrmann, M. C., Hess, M. H., Johns, O.,  Lamppa, D. C., Martin, M. R., McBride, R. D., Peterson, K. J., Porter, J. L., Robertson, G. K., Rochau, G. A., Ruiz, C. L., Savage, M. E., Smith, I. C., Stygar, W. A., and Vesey, R. A.} (2014). Experimental Demonstration of Fusion-Relevant Conditions in Magnetized
Liner Inertial Fusion, PRL 113, 155003 \\
\bibitem[\protect\citeauthoryear{Hohenberger et al.} {2012}]{hohenburger} {\bf Hohenberger, M., Chang, P.-Y., Fiksel, G., Knauer, J. P., Betti, R.,
Marshall, F. J., Meyerhofer, D. D., Séguin, F. H., and Petrasso, R. D.} (2012). Inertial confinement fusion implosions with imposed magnetic field compression using the OMEGA Laser, Phys. Plasmas 19, 056306.\\	
\bibitem[\protect\citeauthoryear{Hora} {2013}]{hora}{\bf Hora, H.} (2013). Extraordinary strong jump of increasing laser fusion gains experienced at volume 
ignition for combination with NIF experiments, Laser and Particle Beams 31, 229 \\
\bibitem[\protect\citeauthoryear{Hurricane et al.} {2014}]{hurricane} {\bf Hurricane, O.A., Callahan, D.A., Casey, D.T., Celliers, P.M., Charles Cerjan, Dewald, E.L., Dittrich, T.R., Döppner, T., Hinkel, D.E., Berzak Hopkins, L.F., Kline, J.L., Le Pape, S., Ma, T., MacPhee, A.G., Milovich, J.L., Pak, A., Park, H-S., Patel, P.K., Remington, B.A., Salmonson, J.D., Springer, P.T. and Tommasini, R.} (2014). Fuel gain exceeding
unity in an inertially confined fusion implosion,Nature 506, 343-349.\\
\bibitem[\protect\citeauthoryear{Kasotakis et al.} {1989}]{kasotakis} {\bf Kasotakis, G., Cicchitelli, L., Hora, H. and Stening, R.J.} (1989). Volume compression and volume ignition of laser driven fusion pellets, 
Laser \& Particle Beams 7, 511.\\
\bibitem[\protect\citeauthoryear{Landau et al.} {1987}] {landau}{\bf Landau, L. D. and Lifshitz, E. M.} (1987). Fluid Mechanics (Pergamon Press)\\
\bibitem[\protect\citeauthoryear{Mallick \& Schramm} {2014}]{mallick} Mallick, R. and Schramm, S. (2014). Oblique magnetohydrodynamic shocks: Space-like and time-like characteristics, Phys. Rev. C 89, 025801. \\
\bibitem[\protect\citeauthoryear{Park et al.} {2014}]{park} {\bf Park, H-S., Hurricane, O. Callahan, A. D.,  Casey, T., Dewald, E. L.,  Dittrich, T. R.,  Döppner, T.,  Hinkel, D. E.,
Berzak Hopkins, L. F.,  Le Pape, S.,  Ma, T. ,  Patel, P. K., Remington, B. A.,  Robey, H. F. and  Salmonson, J. D.} (2014). High-Adiabat High-Foot Inertial Confinement Fusion Implosion Experiments on the National Ignition Facility, Phys. Rev. Lett. 112, 055001.\\
\bibitem[\protect\citeauthoryear{Slutz et al.} {2010}]{slutz} {\bf  Slutz, S. A., Herrmann, M. C.,  Vesey, R. A.,  Sefkow, A. B.,  Sinars, D. B., Rovang, D. C., Peterson, K. J., and Cuneo, M. E.} (2010). Pulsed-power-driven cylindrical liner implosions of laser preheated fuel magnetized with an axial field , Phys. Plasmas 17, 056303. \\	
\bibitem[\protect\citeauthoryear{Taub} {1948}]{taub} {\bf Taub, A. H.}(1948). Relativistic Rankine-Hugoniot Equations, Phys. Rev. {bf 74}, 328 \\
\bibitem[\protect\citeauthoryear{Wasserman} {2011}]{wasserman} {\bf Wasserman, A. L.} (2011). Thermal Physics: Concept and Practice (Cambridge University Press)

\end{thebibliography}
\end{document}